\documentclass[conference]{IEEEtran}
\IEEEoverridecommandlockouts
\usepackage{cite}
\usepackage{amsmath,amssymb,amsfonts}
\usepackage{algorithmic}
\usepackage{graphicx}
\usepackage{textcomp}
\usepackage{xcolor}
\usepackage{multirow}
\usepackage{multicol}
\usepackage{xurl}
\def\BibTeX{{\rm B\kern-.05em{\sc i\kern-.025em b}\kern-.08em
    T\kern-.1667em\lower.7ex\hbox{E}\kern-.125emX}}
\usepackage{subcaption}
\usepackage{float}

\begin{document}
\title{StrikeWatch: Wrist-worn Gait Recognition with Compact Time-series Models on Low-power FPGAs\thanks{The authors gratefully acknowledge the financial support provided by the Federal Ministry for Economic Affairs and Climate Action of Germany for the RIWWER project (01MD22007C).}}
\author{
Tianheng Ling, 
Chao Qian,
Peter Zdankin, 
Torben Weis and Gregor Schiele \\
\IEEEauthorblockA{University of Duisburg-Essen, Duisburg, Germany}
\IEEEauthorblockA{PALUNO, The Ruhr Institute for Software Technology, Essen, Germany}
\IEEEauthorblockA{Email: firstname.lastname@uni-due.de}}
\maketitle
\begin{abstract}
Running offers substantial health benefits, but improper gait patterns can lead to injuries, particularly without expert feedback. While prior gait analysis systems based on cameras, insoles, or body-mounted sensors have demonstrated effectiveness, they are often bulky and limited to offline, post-run analysis. Wrist-worn wearables offer a more practical and non-intrusive alternative, yet enabling real-time gait recognition on such devices remains challenging due to noisy Inertial Measurement Unit (IMU) signals, limited computing resources, and dependence on cloud connectivity.
This paper introduces \emph{StrikeWatch}, a compact wrist-worn system that performs entirely on-device, real-time gait recognition using IMU signals. As a case study, we target the detection of heel versus forefoot strikes to enable runners to self-correct harmful gait patterns through visual and auditory feedback during running. We propose four compact DL architectures (1D-CNN, 1D-SepCNN, LSTM, and Transformer) and optimize them for energy-efficient inference on two representative embedded Field-Programmable Gate Arrays (FPGAs): the AMD Spartan-7 XC7S15 and the Lattice iCE40UP5K.
Using our custom-built hardware prototype, we collect a labeled dataset from outdoor running sessions and evaluate all models via a fully automated deployment pipeline. Our results reveal clear trade-offs between model complexity and hardware efficiency. Evaluated across 12 participants, 6-bit quantized 1D-SepCNN achieves the highest average F1 score of 0.847 while consuming just 0.350 $\mu$J per inference with a latency of 0.140 ms on the iCE40UP5K running at 20 MHz. This configuration supports up to 13.6 days of continuous inference on a 320 mAh battery.
All datasets and code are available in the GitHub repository\footnote{\url{https://github.com/tianheng-ling/StrikeWatch}}. 
\end{abstract}
\begin{IEEEkeywords}
Wrist-Worn Wearables, 
Running Gait Recognition, 
Time-Series Models, 
Model Quantization, 
On-Device Inference, 
Low-Power FPGA
\end{IEEEkeywords}


\section{Introduction}
\label{sec:introduction}

Running is one of the most widely practiced sports worldwide, offering significant physical and mental benefits~\cite{benson2022real}. However, without professional guidance, many runners unknowingly adopt improper gait patterns, such as excessive heel striking that increases the risk of acute and chronic injuries~\cite{burke2021risk}. Although experienced coaches can provide on-side gait correction, such personalized supervision is often expensive and inaccessible to casual or recreational runners.

To bridge this gap, researchers have explored various sensing technologies for gait analysis without requiring direct human supervision. These include image-based systems~\cite{karakasis2021f}, pressure-sensitive insoles~\cite{schiewe2020study, hassan2017footstriker}, and body-mounted \emph{Inertial Measurement Units} (IMUs) placed on the legs or shoes~\cite{muhamad2023design, Young2023, Zago2021}. While effective in controlled environments, these methods typically involve bulky setups and are often limited to offline, post-run analysis. They can not offer immediate corrective feedback during running training, particularly in outdoor environments for long-term usage.

The widespread adoption of wrist-worn wearables such as smartwatches presents a more practical and less obtrusive alternative for continuous gait monitoring~\cite{Kandpal2023}. These devices are compact, widely available, and typically equipped with built-in IMUs. However, leveraging wrist-mounted IMUs to detect fine-grained gait events (such as footstrike types) remains challenging due to lower signal fidelity and confounding arm dynamics~\cite{cola2017personalized}. Moreover, the constrained computing and energy budgets of wearable devices hinder the deployment of advanced \emph{Deep Learning} (DL) time-series models. While offloading to the cloud or edge servers is a possible workaround, it introduces latency, raises privacy concerns, and becomes unreliable in outdoor settings with poor connectivity.

\begin{figure}[!htbp]
\vspace{-10pt}
\centering
    \begin{minipage}[t]{0.4\columnwidth}
        \centering
        \includegraphics{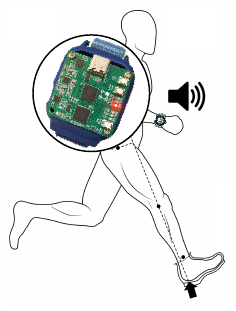}
        \subcaption{Heel striking}
    \end{minipage}
    \begin{minipage}[t]{0.4\columnwidth}
        \centering
        \includegraphics{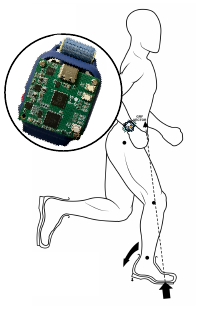}
        \subcaption{Forefoot striking}
    \end{minipage}
     \caption{Wrist-worn \emph{StrikeWatch} system for real-time heel strike recognition, providing on-device visual (LED) and auditory feedback. Runner silhouettes adapted from~\cite{RunningImage}.}
    \label{fig:strike_watch}
\vspace{-5pt}
\end{figure}

To address these limitations while enabling autonomous edge intelligence for IoT, this study proposes \emph{StrikeWatch}, a wrist-worn system that integrates software-hardware co-design for on-device gait recognition in practical outdoor settings (see Figure~\ref{fig:strike_watch}). As a case study, we classify forefoot versus heel strikes from wrist-mounted IMU signals, enabling runners to self-correct harmful gait patterns through real-time visual and auditory feedback provided onboard.
More in detail, key contributions are summarized as follows:
\begin{itemize}
    \item We develop a 32-gram hardware prototype of the \emph{StrikeWatch} system that tightly integrates sensing, computing, and feedback in a watch-sized wrist-worn form factor. By enabling real-time inference on low-power \emph{Field-Programmable Gate Arrays} (FPGAs), our design demonstrates a practical solution for embedded gait recognition.

    \item We design and optimize four lightweight DL models (1D-CNN, 1D-SepCNN, LSTM, and Transformer) for running gait recognition on two representative resource-constrained embedded FPGAs (AMD Spartan-7 XC7S15 and Lattice iCE40UP5K).

    \item Through a fully automated deployment pipeline, we evaluate all models on real hardware and report empirical metrics, including accuracy, latency, power, and energy. These results provide a comparative foundation for understanding the trade-offs between model complexity and efficiency in resource-constrained edge AI systems.
    
    \item We contribute labeled datasets collected from real-world outdoor running sessions and release all code and implementation details to promote reproducibility and foster future research on embedded wearables.
    
    
\end{itemize}

The remainder of this paper is organized as follows:  
Section~\ref{sec:related_work} reviews prior work on IMU-based gait recognition.  
Section~\ref{sec:strikewatch_prototype} presents the hardware design of the \emph{StrikeWatch} prototype.  
Section~\ref{sec:data_collection} details the data acquisition and preprocessing process. 
Section~\ref{sec:model_architecture} introduces the model architectures.  
Section~\ref{sec:deloyment_pipeline} describes the deployment pipeline with optimizations for resource-constrained FPGAs.  
Section~\ref{sec:feedback} outlines the feedback trigger mechanism.
Section~\ref{sec:experiemnt_results} presents experimental results and evaluation.  
Finally, Section~\ref{sec:conclusion_future_work} concludes the paper and outlines future work.

\section{Related Work}
\label{sec:related_work}

\begin{table*}[!htb]
\footnotesize
\centering

\setlength{\tabcolsep}{.4mm}{
\caption{Comparison of IMU-based running gait recognition studies}
\label{tab:imu_comparison}
\begin{tabular}{|c|ccccccccccc|}
\hline
\multicolumn{1}{|c|}{Description} & \multicolumn{1}{c|}{Yuwono et al.~\cite{Yuwono2013}} & \multicolumn{1}{c|}{Mahoney et al.~\cite{Mahoney2024}} & \multicolumn{6}{c|}{Joo et al.~\cite{Joo2022}} & \multicolumn{1}{c|}{Giandolini et al.~\cite{Giandolini2014}} & \multicolumn{1}{c|}{Young et al.~\cite{Young2023}} & \multicolumn{1}{c|}{Zago et al.~\cite{Zago2021}} \\ \hline
\begin{tabular}[c]{@{}c@{}}Recognition\\Pattern\end{tabular}  & \multicolumn{1}{c|}{H} & \multicolumn{1}{c|}{H, M, F}  & \multicolumn{6}{c|}{H, M, F} & \multicolumn{1}{c|}{H, M, F} & \multicolumn{1}{c|}{H, HM, M, MF, F} & \multicolumn{1}{c|}{H} \\ \hline

\begin{tabular}[c]{@{}c@{}}Sensor \\ Placement \end{tabular} & \multicolumn{1}{c|}{Waist} & \multicolumn{1}{c|}{Ankle}  & \multicolumn{6}{c|}{Wrist} & \multicolumn{3}{c|}{Shoes} \\ \hline

\begin{tabular}[c]{@{}c@{}}Sampling\\ Frequency(Hz)\end{tabular} & \multicolumn{1}{c|}{50} & \multicolumn{1}{c|}{1000}  & \multicolumn{6}{c|}{50} & \multicolumn{1}{c|}{1000} & \multicolumn{1}{c|}{60} & \multicolumn{1}{c|}{512}  \\ \hline

\begin{tabular}[c]{@{}c@{}}Input \\ Features\end{tabular} & \multicolumn{1}{c|}{$a_y$, $a_z$} & \multicolumn{7}{c|}{$a_x$, $a_y$, $a_z$} & \multicolumn{1}{c|}{2 uniaxial} & \multicolumn{1}{c|}{3+3} & \multicolumn{1}{c|}{3+3} \\ \hline 

\begin{tabular}[c]{@{}c@{}} Window \\ Size \end{tabular} & \multicolumn{1}{c|}{66} & \multicolumn{1}{c|}{40, 75, 100}  & \multicolumn{6}{c|}{50, 75, 100, 150} & \multicolumn{2}{c|}{-} & \multicolumn{1}{c|}{64} \\ \hline
\begin{tabular}[c]{@{}c@{}} Algorithm \end{tabular} & \multicolumn{1}{c|}{\begin{tabular}[c]{@{}c@{}}Feature \\ Extraction+ \\ Clustering\end{tabular}} & \multicolumn{1}{c|}{\begin{tabular}[c]{@{}c@{}}ANN+\\Voting\end{tabular}}  & \multicolumn{1}{c|}{\begin{tabular}[c]{@{}c@{}}Naive \\ Bayes\end{tabular}} & \multicolumn{1}{c|}{\begin{tabular}[c]{@{}c@{}}Random \\ Forest\end{tabular}} & \multicolumn{1}{c|}{SVM} & \multicolumn{1}{c|}{LSTM} & \multicolumn{1}{c|}{GRU} & \multicolumn{1}{c|}{CNN} & \multicolumn{1}{c|}{\begin{tabular}[c]{@{}c@{}}Temporal Detection\\ of Peak Accelerations\end{tabular}} & \multicolumn{1}{c|}{\begin{tabular}[c]{@{}c@{}}Fuzzy \\ Logic \end{tabular}} &  \multicolumn{1}{c|}{\begin{tabular}[c]{@{}c@{}}Decision \\ Tree\end{tabular}} \\ \hline

\begin{tabular}[c]{@{}c@{}}Num. of \\ Particitants\end{tabular}  & \multicolumn{1}{c|}{8} & \multicolumn{1}{c|}{58}  & \multicolumn{6}{c|}{17} & \multicolumn{1}{c|}{34} & \multicolumn{1}{c|}{203} & \multicolumn{1}{c|}{40} \\ \hline 

\begin{tabular}[c]{@{}c@{}}Training\\ Approach \end{tabular} & \multicolumn{1}{c|}{{I, GA}} & \multicolumn{1}{c|}{{GL2}}&\multicolumn{6}{c|}{{GA}} &\multicolumn{1}{c|}{{GC}}&\multicolumn{1}{c|}{{GR10Fold}} &\multicolumn{1}{c|}{{GR}}  \\ \hline

\begin{tabular}[c]{@{}c@{}}Metrics \end{tabular} & \multicolumn{1}{c|}{{Accuracy(\%)}} &\multicolumn{1}{c|}{{Accuracy(\%), F1 score}} &  \multicolumn{6}{c|}{F1 score} & \multicolumn{1}{c|}{CC$^{\star}$} & \multicolumn{1}{c|}{ICC$^{\star}$}   & \multicolumn{1}{c|}{Accuracy(\%)} \\ \hline

\begin{tabular}[c]{@{}c@{}}Results \\ (up to)\end{tabular}  & \multicolumn{1}{c|}{{96.7, 86.7}} & \multicolumn{1}{c|}{93.3, 0.899}  &  \multicolumn{1}{c|}{0.561} & \multicolumn{1}{c|}{0.780} & \multicolumn{1}{c|}{0.953} & \multicolumn{1}{c|}{0.968} & \multicolumn{1}{c|}{0.974} & \multicolumn{1}{c|}{0.980} & \multicolumn{1}{c|}{0.946} & \multicolumn{1}{c|}{0.90}  & \multicolumn{1}{c|}{93.3} \\ \hline
\begin{tabular}[c]{@{}c@{}}Device\end{tabular} & \multicolumn{1}{c|}{Customized} & \multicolumn{1}{c|}{Customized}  & \multicolumn{6}{c|}{Sport Smartwatch FTW6024 by Fossil} &\multicolumn{1}{c|}{Customized }&\multicolumn{1}{c|}{Mymo Sensor}&\multicolumn{1}{c|}{Physilog} \\ \hline

\begin{tabular}[c]{@{}c@{}} Inference \\ Time(ms)\end{tabular}  & \multicolumn{2}{c|}{-} & \multicolumn{1}{c|}{3.03} & \multicolumn{1}{c|}{9.38} & \multicolumn{1}{c|}{3.7} & \multicolumn{1}{c|}{35.9} & \multicolumn{1}{c|}{36.3} & \multicolumn{1}{c|}{27.8}& \multicolumn{3}{c|}{-} \\ \hline

\multicolumn{12}{l}{$^{\star}$ CC refers to correlation coefficients, and ICC to intraclass correlation coefficients. Neither study~\cite{Giandolini2014,Young2023} reports classification accuracy.}\\
\end{tabular}}
\vspace{-10pt}
\end{table*}

IMU-based methods have emerged as a practical choice for running gait recognition due to their portability and ability to capture rich motion dynamics. In this section, we review prior work from three perspectives: (1) sensing configuration and signal processing, (2) model architectures and training strategies, and (3) system deployment on embedded hardware. Table~\ref{tab:imu_comparison} provides a summary of representative studies.

\subsection{IMU-based Running Gait Recognition}

Prior studies target multiple gait patterns, such as heel (H), mid-foot (M), forefoot (F), or their combinations (e.g., HM or MF). For deployment simplicity, we focus on recognizing two primary patterns (heel strike and forefoot strike), which are critical for assessing injury risk and guiding correction.

Sensor placement significantly affects both the quality and practicality of gait signal acquisition. Most prior works place IMUs near the lower body, such as on the ankle~\cite{Mahoney2024}, shoes~\cite{Young2023, Giandolini2014, Zago2021}, or waist~\cite{Yuwono2013}, to obtain high-fidelity signals. However, these placements are often bulky and intrusive, limiting their suitability for daily or long-term real-world use. Motivated by~\cite{Joo2022}, we adopt a wrist-worn IMU in a watch form factor to enhance wearability and accessibility. Although wrist placement introduces greater signal noise and arm motion artifacts, model adoption can mitigate these limitations.
Given the wrist-worn setup and embedded constraints, sampling frequency must balance signal resolution and energy cost. Prior studies have explored frequencies ranging from 50 Hz~\cite{Yuwono2013, Joo2022} to 1000 Hz~\cite{Mahoney2024, Giandolini2014}. We select 100 Hz as a balanced setting that provides sufficient temporal resolution while remaining efficient for on-device processing.

Feature selection also varies considerably across studies, depending on sensor placement and preprocessing strategy. Yuwono et al.~\cite{Yuwono2013} used biaxial accelerometer readings (\(a_y\), \(a_z\)) from a waist-mounted sensor. In contrast, Joo et al.~\cite{Joo2022} and Mahoney et al.~\cite{Mahoney2024} adopted full triaxial inputs (\(a_x\), \(a_y\), \(a_z\)) to capture more comprehensive motion dynamics. Giandolini et al.~\cite{Giandolini2014} used two uniaxial sensors placed directly on the feet. Young et al.~\cite{Young2023} and Zago et al.~\cite{Zago2021} further combined accelerometer and gyroscope data to enhance rotational awareness. In our case, we systematically compared raw triaxial acceleration and derived features, such as acceleration magnitude ($a\!=\!\sqrt{a_x^2 + a_y^2 + a_z^2}$). While derived features occasionally improved individual performance, they did not yield consistent gains across all participants and sometimes introduced noise. Thus, we retain raw triaxial acceleration as input.
The input sequence length must also be chosen with care to model wrist-based IMU signals. Prior studies have used input lengths ranging from 40 to 150 samples~\cite{Joo2022, Mahoney2024}. Based on our preliminary experiments, we selected 50 samples as input (i.e., 0.5 second at 100 Hz), which captures sufficient motion context while remaining compatible with the resource constraints of embedded FPGAs.

\subsection{Algorithms for Running Gait Recognition}

Running gait recognition has evolved from traditional signal processing to modern DL techniques. Early approaches, such as those by Yuwono et al.~\cite{Yuwono2013}, relied on handcrafted features extracted from raw IMU data. While lightweight, such methods required domain expertise and often failed to generalize to diverse gait patterns. To improve robustness, later studies adopted statistical techniques such as peak acceleration detection~\cite{Giandolini2014} and fuzzy logic systems~\cite{Young2023}, which improved interpretability but still struggled with subject variability and unseen conditions.

The introduction of \emph{Machine Learning} (ML) enabled more adaptive models. Mahoney et al.~\cite{Mahoney2024} proposed an \emph{Artificial Neural Network} (ANN) combined with a voting mechanism to stabilize predictions, while Zago et al.~\cite{Zago2021} employed \emph{Decision Trees}. Joo et al.~\cite{Joo2022} evaluated various ML algorithms, including \emph{Naive Bayes}, \emph{Random Forest}, and \emph{Support Vector Machines} (SVM), showing improved accuracy over traditional techniques. However, these ML models cannot capture temporal dependencies in gait signals.

DL methods offer stronger modeling capacity by directly learning from raw IMU sequences~\cite{Joo2022}. 1D-\emph{Convolutional Neural Networks} (1D-CNNs) have been widely used to extract local temporal patterns, while \emph{Long Short-Term Memory} (LSTM) networks are capable of modeling long-range dependencies. More recently, Transformer-based models have shown promise for time-series tasks due to their self-attention mechanism, which enables parallel sequence modeling and faster convergence~\cite{ling2024aiot}. Motivated by these advances, we explore and optimize four representative DL architectures (1D-CNN, 1D-SepCNN, LSTM, and Transformer) for running gait recognition on resource-constrained hardware.

Prior studies also differ in training strategies and are summarized as follows: Individualized training (I), where models are tailored per subject; Generalized training by averaging across participants (GA); Generalized leave-two-subject-out cross-validation (GL2); Generalized training with separate training and validation datasets (GS); Generalized 10-fold cross-validation with random splits (GR10Fold); and Generalized training via random splits without subject separation (GR). 

We initially compared three training strategies: individualized training, generalized training with leave-one-subject-out evaluation, and a fine-tuning strategy that adapts the generalized model using a small amount of subject-specific data. Among these three, fine-tuning offered the best trade-off between accuracy and generalization, with lower inter-subject variance and minimal user-specific data requirements. For example, using the Transformer-based model in floating-point format, the individualized strategy achieved an average best F1 score of 0.925 (std 0.054) across all participants, while the generalized strategy dropped to 0.701 (std 0.191) due to large inter-subject variability. The fine-tuning strategy bridged this gap by improving performance to 0.952 (std 0.041). Therefore, we adopt it as the default strategy in subsequent experiments.

\subsection{On-device Running Gait Recognition}

Studies shown in Table \ref{tab:imu_comparison} collected data using custom or commercial devices, with both training and inference offloaded to external high-performance computers. In contrast, our system executes inference directly on the wearable, enabling real-time gait recognition under tight resource constraints.
Recent work~\cite{ling2024aiot} demonstrated that executing Transformer models on MCUs~\cite{becnel2022tiny} is often less efficient than using embedded FPGAs. Motivated by this, we adopt a heterogeneous architecture where an embedded FPGA supplements the MCU to accelerate model inference.

To understand the practical viability of such FPGA-based acceleration in wearable scenarios, we reviewed existing platforms and their power-performance characteristics. Several prior studies have explored FPGA-based DL model acceleration for wearable or embedded systems. Roggen et al.~\cite{roggen2022wearable} proposed a wearable platform using the Intel MAX10 FPGA for digital signal processing tasks such as filtering. However, its power consumption exceeds 300 mW, making it unsuitable for always-on wearable use.
Qian et al.\cite{qian2023elasticai} introduced an end-to-end deployment toolchain for DL models on AMD Spartan-7 XC7S15, demonstrating that LSTM models can be deployed successfully, but with power consumption exceeding 70 mW. In addition, Li et al.~\cite{li2024hardware} deployed 1D-CNN models on the XC7S15 for radar-based gesture recognition, confirming its capability for high-throughput inference but also noting its relatively high power consumption.

Moreover, Chen et al.\cite{chen2024eciton} compared LSTM accelerators on the XC7S15 and the smaller Lattice iCE40UP5K FPGA, showing that while the XC7S15 offers higher throughput, the iCE40UP5K is significantly more power-efficient, benefiting from sub-milliwatt static power. A broader survey by Chen et al.\cite{chen2021mlof} across ten FPGAs further confirms that the iCE40UP5K offers the best power–performance trade-off for lightweight DL inference.
Motivated by these findings, we adopt both the AMD XC7S15 and Lattice iCE40UP5K as target platforms for our \emph{StrikeWatch} deployment. These two FPGAs span different points in the power–performance spectrum, allowing us to systematically investigate trade-offs between energy efficiency and computational capability.

\section{Hardware Prototype}
\label{sec:strikewatch_prototype}

This section introduces the hardware prototype of \emph{StrikeWatch}, a compact wrist-worn device designed to perform on-device DL inference under stringent size and energy constraints. Unlike previous bulky or cloud-dependent systems, \emph{StrikeWatch} integrates sensing, computing, and feedback into a fully self-contained, low-latency wearable.

\begin{figure}[!htbp]
\vspace{-5pt} 
    \centering
    \begin{minipage}[t]{0.46\columnwidth}
        \centering
        \includegraphics[width=\textwidth]{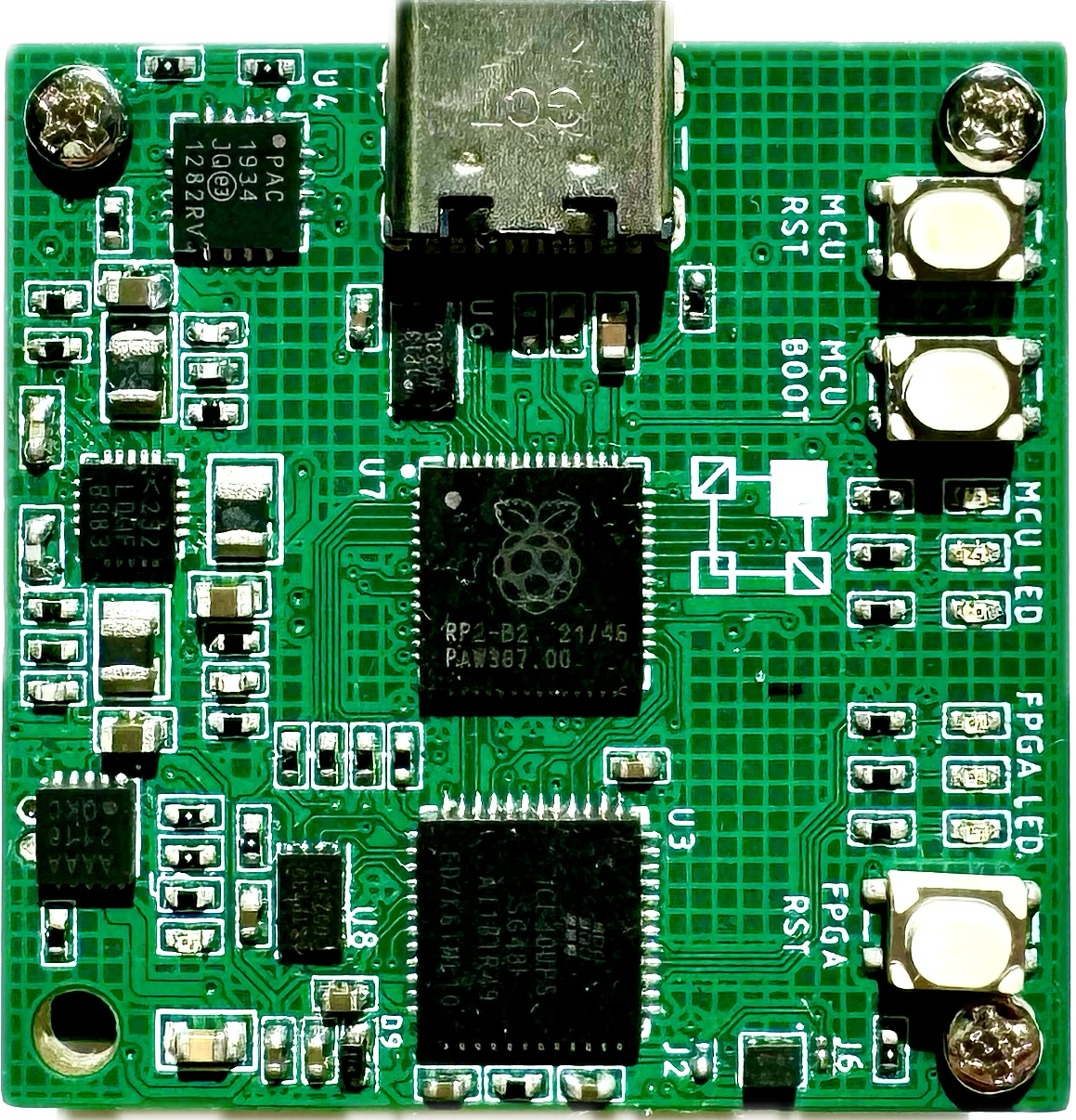}
        \centerline{(1) Compute board}
    \end{minipage}
    \begin{minipage}[t]{0.48\columnwidth}
        \centering
        \includegraphics[width=\textwidth, angle=90]{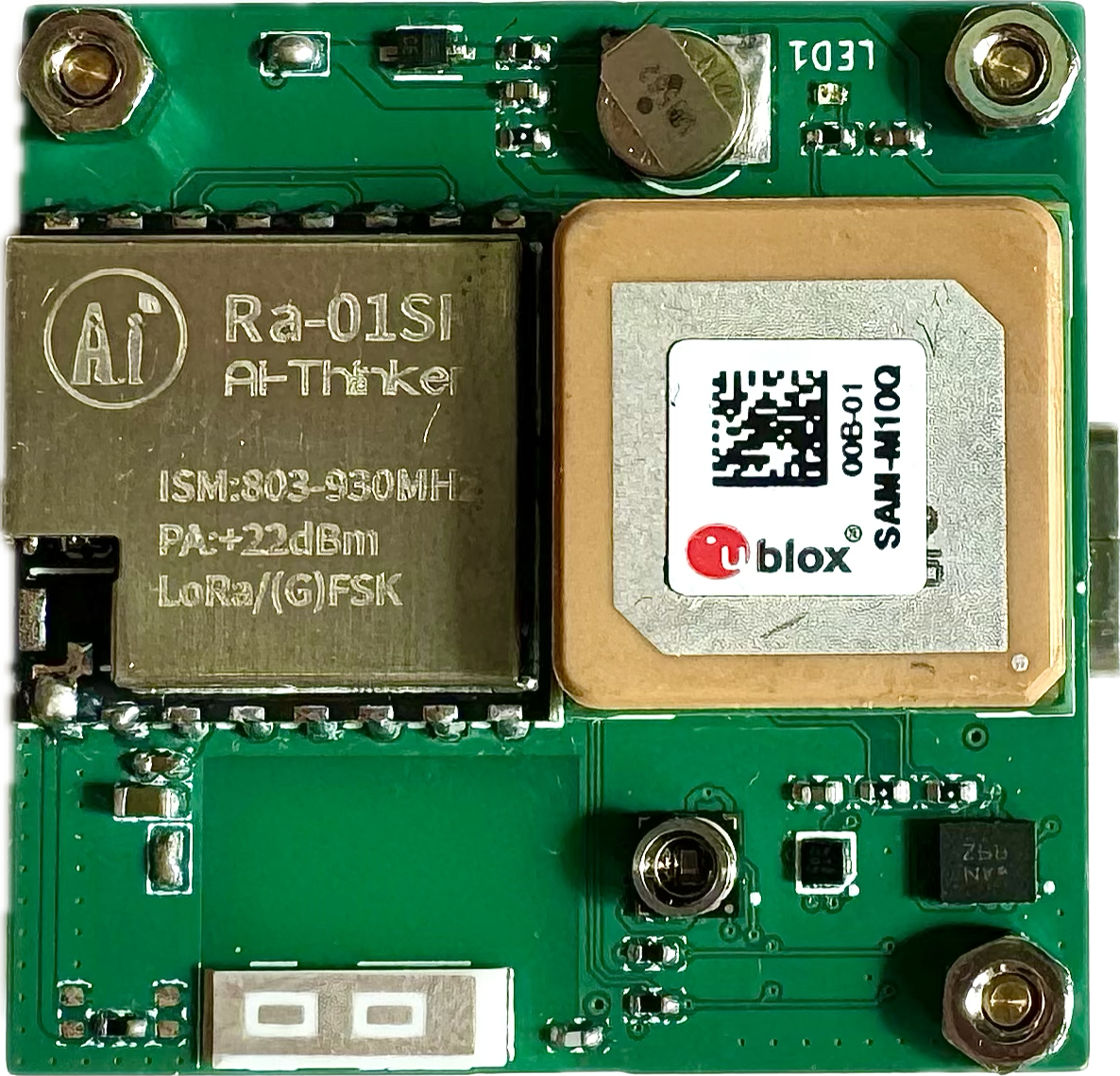}
        \centerline{(2) Application board}
    \end{minipage}
    
    \vspace{5pt}
    
    \begin{minipage}[t]{\columnwidth}
        \centering
        \includegraphics[width=.95\textwidth]{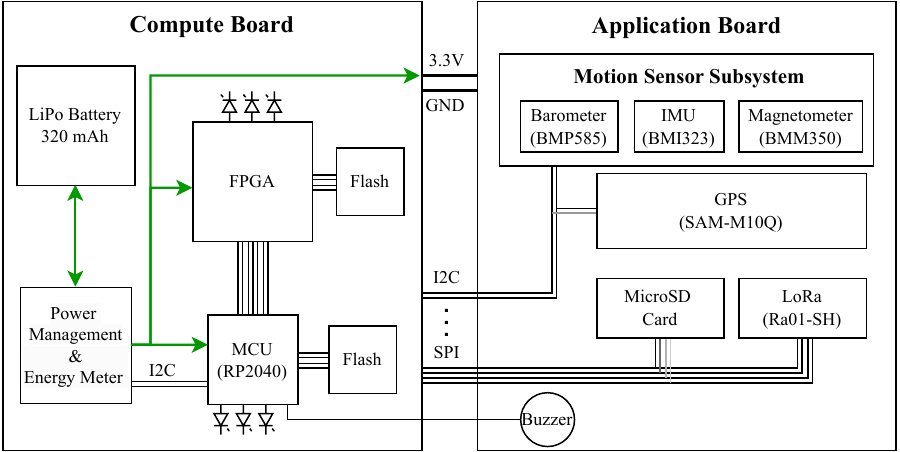}
        \centerline{(3) Schematic diagram}
    \end{minipage}
    \caption{\emph{StrikeWatch} hardware and schematic diagram}
    \label{fig:hardware}
\vspace{-5pt}    
\end{figure}

As shown in Figure~\ref{fig:hardware}, the hardware measures $34\times34\times17$~mm and weighs 32~gram, making it suitable for continuous use during running. The prototype is split into two tightly integrated boards to separate core inference tasks from sensing and feedback. 
The \emph{compute board} handles DL inference and power management. It features an RP2040 MCU responsible for system control, sensor sampling, and SPI/I2C communication. The model inference is offloaded to an embedded FPGA, supporting both the AMD XC7S15 and the Lattice iCE40UP5K. A 320~mAh LiPo battery powers the device, while an onboard energy meter allows fine-grained measurement of compute and sensing components to support realistic power evaluation. 
The \emph{application board} is tailored for sensing, storage, and communication. It includes a BMI323 IMU for capturing wrist motion and a microSD card for recording raw sensor logs. Additional modules such as a BMP585 barometer, a SAM-M10Q GPS module, and a Ra01-SH LoRa transceiver are also included for future use cases like environmental awareness or remote data synchronization. 
Feedback is delivered via an onboard buzzer and LED indicator, enabling immediate auditory and visual cues during running. The two boards communicate via standard I2C and SPI buses: the IMU connects to the MCU over I2C, while SPI is used for model I/O and data logging. 


\section{Data Collection and Preprocessing}
\label{sec:data_collection}

For data collection, we recruited 16 participants (3 female, 13 male, aged 25–34, weighing 55.3–114 kg), each completing two 1-minute outdoor running sessions while wearing \emph{StrikeWatch}. As shown in Figure \ref{fig:runing_gait}, participants were instructed to adopt a forefoot strike gait during the first session and a heel strike gait during the second, enabling within-subject comparison of gait types. The short session duration is chosen to minimize potential strain on participants. Wrist motion signals reflecting different running gaits were captured by the BMI323 IMU (100 Hz, ±2$g$) and stored on a microSD card.

All participants were instructed to maintain a consistent arm-swing pattern during both sessions to ensure reliable gait differentiation from wrist data. An examiner followed each participant in parallel during running and recorded their running sessions using an iPhone 15 Pro (60 FPS). These video recordings were later synchronized with the collected IMU data and used to manually label each step as \textit{forefoot strike} or \textit{heel strike}. Each participant’s data consists of the IMU measurements stored in JSON format and corresponding video files, providing ground truth for supervised training.

\noindent
\begin{minipage}{\columnwidth}
    \begin{figure}[H]
        \centering
        \begin{minipage}[t]{0.49\textwidth}
            \centering
            \includegraphics[height=140pt]{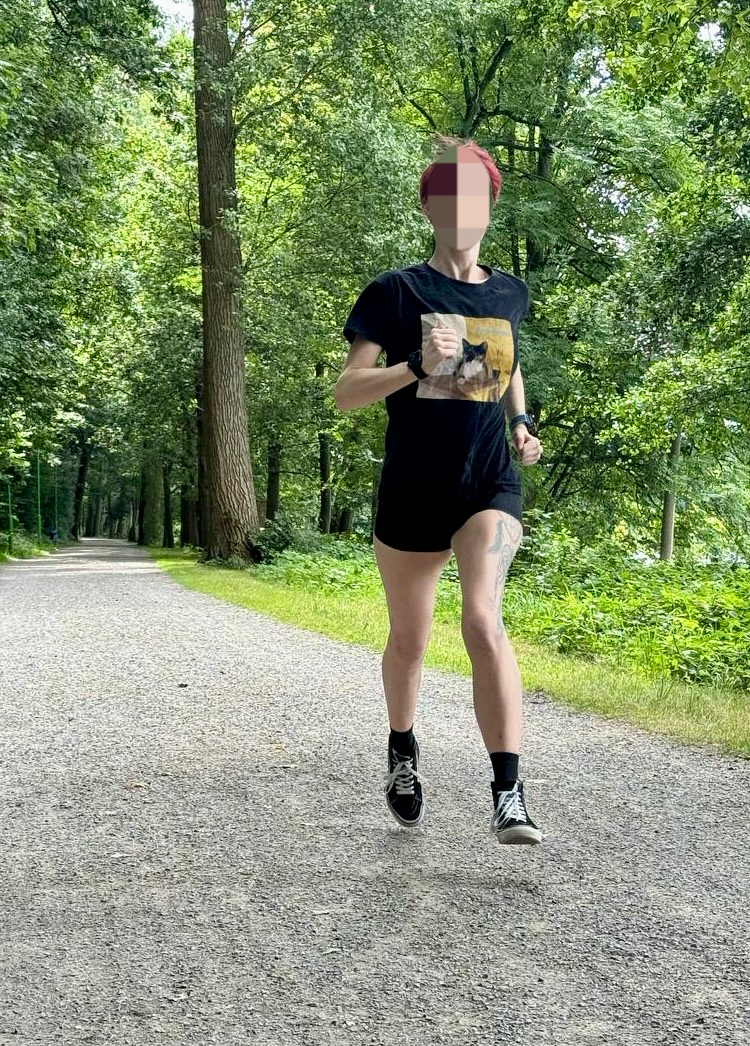}
        \end{minipage}%
        \hfill
        \begin{minipage}[t]{0.49\textwidth}
            \centering
            \includegraphics[height=140pt]{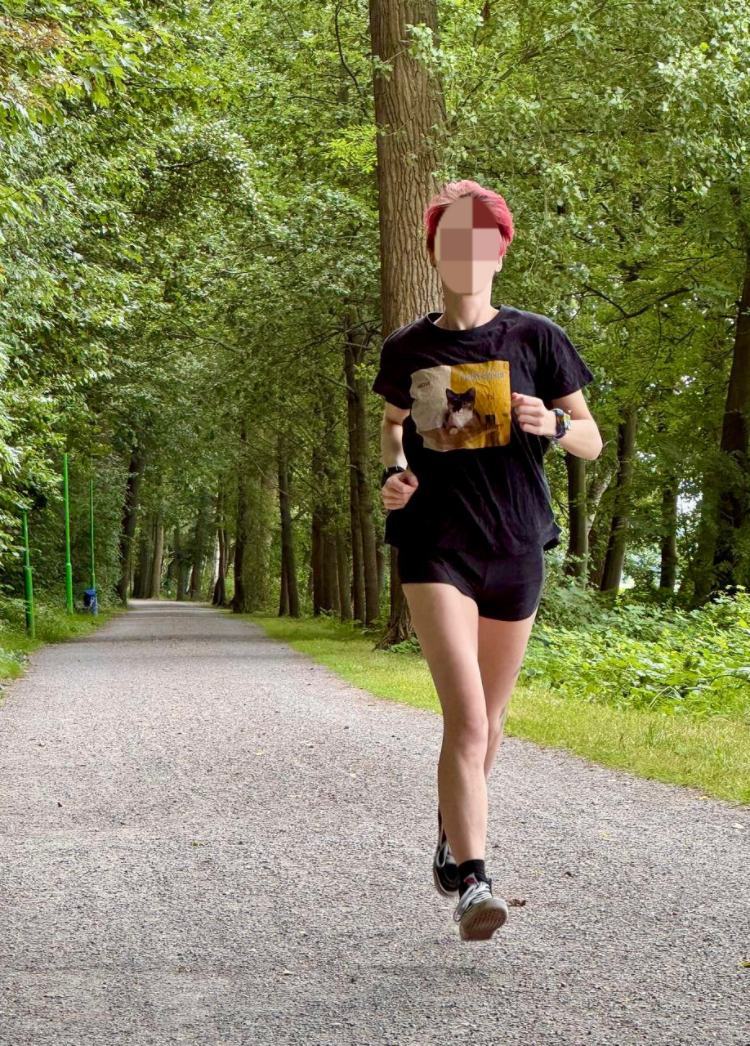}
        \end{minipage}
        \caption{Example of forefoot (left) and heel (right) strikes, captured from a participant during outdoor data collection.}
        \label{fig:runing_gait}
    \end{figure}
    \vspace{-15pt}
    \begin{figure}[H]
        \centering
        \includegraphics[width=\textwidth]{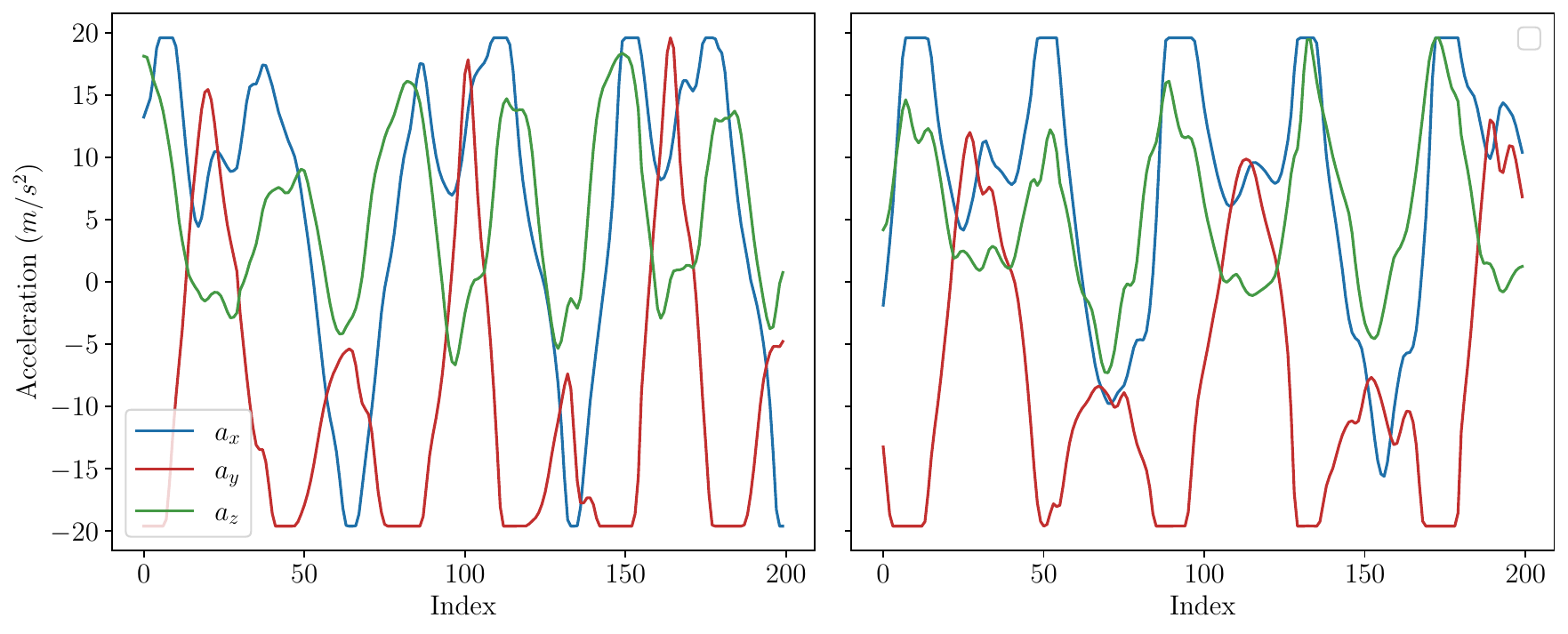}
        \caption{Sample triaxial wrist-worn acceleration signals for forefoot (left) and heel (right) strike instances.}
        \label{fig:runing_gait_signal}
    \end{figure}
 \vspace{-5pt}
\end{minipage}

A manual video review confirmed that most participants followed the instructed gait patterns. However, two participants (one female and one male) displayed atypical gait patterns that could not be reliably classified as either forefoot or heel strikes, potentially due to flat foot conditions reported by them. Since clear ground-truth labels were not available, we excluded all data from these two participants from subsequent experiments. Interestingly, this observation also highlights that certain foot anatomies, such as flat feet, may introduce ambiguity in gait classification based on wrist-mounted IMU signals. In addition, two further participants were excluded due to incomplete video recordings.
For the remaining 12 participants, we removed transition periods at the beginning and end of each session and walking segments between the two sessions. We then segmented the cleaned time series into the forefoot and heel strike categories. To ensure class and participant balance, we capped each gait type at 3,800 data points per participant, using the minimum count across the dataset and discarding excess data to maintain uniformity. As illustrated in Figure~\ref{fig:runing_gait_signal}, the raw triaxial acceleration signals (\(a_x\), \(a_y\), and \(a_z\)) exhibit clearly distinct patterns between forefoot and heel strike gaits. No additional filtering was applied. 

\section{Model Architectures}
\label{sec:model_architecture}

Taking raw triaxial accelerometer sequences recorded from the wrist as input, we design four compact time-series models.

\textbf{1D-CNN} – Inspired by~\cite{lingaiot2025smatbale,Joo2022}, this model consists of $\text{num}_\text{blocks}$ temporal convolutional blocks. Each block applies a 1D convolution (kernel size 3, stride 1), followed by batch normalization and ReLU activation. To reduce intermediate buffer size, 1D max pooling (kernel size 2) is applied after each block except the last. The first two blocks use narrow output channels of 3 to minimize memory usage. In subsequent blocks, the number of output channels doubles every two blocks (e.g., 6 channels for Block 3 and 4), enhancing representational capacity as temporal resolution decreases. After the final block, a global average pooling layer aggregates features over the temporal dimension. The resulting vector passes through two dense layers: the first applies ReLU activation, and the second produces the final classification logits.

\textbf{1D-SepCNN} – Following the 1D-CNN, this variant replaces standard 1D convolutions with 1D depthwise-separable convolutions to reduce parameters and computational cost. Each block now consists of a depthwise convolution (applying one filter per input channel) followed by a pointwise convolution (kernel size 1) to mix channel-wise features. 

\textbf{LSTM} – This model comprises a vanilla LSTM layer with the number of hidden units ($h_\text{size}$), followed by a dense layer. The final hidden state from the LSTM layer is passed to the dense layer to generate classification logits. To improve FPGA compatibility, we adopt hardware-friendly activation functions: \emph{HardSigmoid} and \emph{HardTanh}, as implemented in~\cite{ling2025automated}. 

\textbf{Transformer} – This model adopts a compact encoder-only architecture inspired by~\cite{ling2024aiot}. It begins with a linear input projection that maps the input into a $d_\text{model}$-dimensional latent space, followed by fixed positional encoding. Next, a single encoder layer applies \emph{One-Head Self-Attention} (OHSA) and a \emph{FeedForward Network} (FFN), each equipped with residual connections and Batch Normalization. The FFN has a hidden dimension of $4 \!\times\! d_\text{model}$ to provide sufficient non-linearity. Final embeddings are aggregated via global average pooling over temporal dimension and passed through a dense layer to produce final logits.

\section{Deployment Pipeline}
\label{sec:deloyment_pipeline}

To enable seamless deployment on embedded FPGAs, we extend the end-to-end pipeline proposed in~\cite{ling2025deployment,lingaiot2025smatbale,ling2025automated} to support the DL models introduced in Section~\ref{sec:model_architecture}.
Each model is constructed using quantizable layers or modules from the \emph{ElasticAI.Creator}\footnote{\url{https://github.com/es-ude/elastic-ai.creator/tree/add-linear-quantization/}} library~\cite{qian2023elasticai}. During training and inference, the library provides two key APIs: \texttt{forward()} for \emph{Quantization-Aware Training} (QAT) down to 4-bit and \texttt{int\_forward()} for integer-only inference using quantized weights and activations.

This library also provides a \texttt{design()} API, which automatically translates all quantized layers into synthesizable RTL using modular VHDL templates, resulting in a complete hardware implementation of the model. For the Transformer architecture, we introduce a customized FFN VHDL template with ping-pong scheduling to reduce intermediate buffer usage. For instance, with $d_{\text{model}}\!=\!16$, 8-bit quantization, and input sequence length of 20, LUT utilization is reduced from 62.18\% to 59.5\% on the AMD XC7S15 FPGA. The generated implementation is validated via RTL simulation using GHDL and synthesized using Vivado (AMD Spartan-7) or Radiant (Lattice iCE40), producing detailed reports on latency, resource usage, and power consumption. To identify model configurations that balance accuracy and energy consumption, the entire process is guided by hardware-aware hyperparameter search using Optuna~\cite{akiba2019optuna,ling2025deployment} within a predefined search space. To establish performance baselines for comparison, each selected configuration is also re-trained using standard full-precision (FP32) layers from PyTorch.

\section{Feedback Trigger Mechanism}
\label{sec:feedback}

Beyond hardware-aware optimization, \emph{StrikeWatch} incorporates a feedback trigger mechanism that balances responsiveness and energy efficiency. Instead of reacting to every isolated prediction, our system issues feedback only when a target gait event is detected across several consecutive inferences. This design mitigates false positives arising from occasional misclassifications. To support this mechanism, we define a set of system-level parameters:
\begin{itemize}
    \item $w$: window size (number of raw samples),
    \item $f$: sampling frequency (Hz),
    \item $s$: stride ratio between adjacent windows ($0 < s \leq 1$),
    \item $d$: temporal downsampling factor,
    \item $n \!=\! \frac{w}{d}$: model input sequence length,
    \item $\text{N}_{\text{consec}}$: number of consecutive positive predictions required to trigger feedback,
    \item $ T_{\text{feedback}} \!=\! (\text{N}_{\text{consec}} - 1) \cdot \frac{w \cdot s}{f}$: minimum feedback latency (excluding the cold-start delay of $\frac{w}{f}$ seconds),
    \item $T_{\text{infer}}$: model inference time per input (in seconds),
    \item $E_{\text{infer}}$: energy consumed per inference (in $\mu$J).
\end{itemize}

This formulation reveals several key design trade-offs. To ensure real-time operation, each inference must be completed before the next input window has been fully accumulated, yielding the constraint $T_{\text{infer}} \leq \tfrac{w \cdot s}{f}$.
A smaller stride $s$ increases the frequency of window updates, thereby reducing $T_{\text{feedback}}$ and improving system responsiveness. However, it also raises the inference frequency, increasing energy consumption per second. In the worst case, the energy consumption rate is upper-bounded by $E_{\text{infer}} \cdot \frac{f}{w \cdot s}$.
The downsampling factor $d$ determines how many samples are retained in each window and thus directly impacts the model input length $n$. A larger $d$ reduces $n$, enabling faster inference (i.e., smaller $T_{\text{infer}}$) and lowering memory and compute requirements. 
Moreover, increasing the threshold $\text{N}_{\text{consec}}$ helps suppress false positives by requiring consistent prediction confidence before issuing feedback. While this improves robustness, it proportionally increases the feedback delay.

In this study, we adopt the following configuration based on empirical analysis: a sampling frequency $f$ of 100~Hz and a window size $w$ of 50, corresponding to 0.5 second of raw IMU data. A stride ratio $s$ of 0.25 results in 75\% overlap between adjacent windows. We set the feedback trigger threshold $\text{N}_{\text{consec}}$ to 5. This configuration yields a minimum feedback latency of 0.5 second, allowing the system to issue up to 2 feedback events per second. To guarantee real-time operation, the model must complete inference within the stride interval, i.e., $T_{\text{infer}} \!<\! 0.125$ second. We apply a downsampling factor of 2 to ensure deployability, empirically determined as the optimal trade-off between model accuracy and computational cost, resulting in an input sequence length of 25.

\section{Experiments and Evaluation}
\label{sec:experiemnt_results}

Building on a complete system design, this section evaluates the effectiveness of \emph{StrikeWatch} through a three-stage experimental study: (1) deployment optimization on a participant, (2) cross-FPGA validation, and (3) across-participant validation.

\subsection{Stage 1: Deployment-Aware Optimization on Participant~1}
\label{subsec:exp1}

Participant 1 is selected as the target subject for hardware-aware optimization at this stage. As described in Section~\ref{sec:related_work}, we adopt a two-step training strategy: (1) leave-one-participant-out training using data from the other 11 participants, and (2) personalized fine-tuning with QAT on the training subset of Participant~1’s data. In both steps, each participant’s data is split into 70\% training, 10\% validation, and 20\% testing.

We conduct 200 trials per model (1D-CNN, 1D-SepCNN, LSTM, and Transformer) using the \texttt{NSGAIISampler}, jointly optimizing for validation F1-score and per-inference energy consumption. All models are trained (on an RTX 2080 SUPER GPU), quantized and synthesized (on a Ryzen Threadripper 3970X CPU), and deployed to an AMD XC7S15 FPGA using the pipeline described in Section~\ref{sec:deloyment_pipeline}. The clock frequency of the XC7S15 FPGA is fixed at 100 MHz.
The search space for the Optuna-based model configuration includes:
\begin{itemize}
    \item Quantization bitwidth: $b \in \{4, 6, 8\}$,
    \item Batch size: $bs \in \{16, 32, \dots, 48\}$,
    \item Learning rate: $lr \in [10^{-5}, 10^{-3}]$ (log-uniform),
    \item 1D-CNN/1D-SepCNN's blocks: $\text{num}_\text{blocks} \in \{1, 2,...,6\}$,
    \item LSTM's hidden size : $h_\text{size} \in \{8, 16, ..., 64\}$,
    \item Transformer's dimension: $d_\text{model} \in \{8, 16, ..., 32\}$.
\end{itemize}
\begin{figure*}[ht]
    \centering
    \begin{subfigure}[b]{0.23\textwidth}
        \centering
        \includegraphics[width=\linewidth]{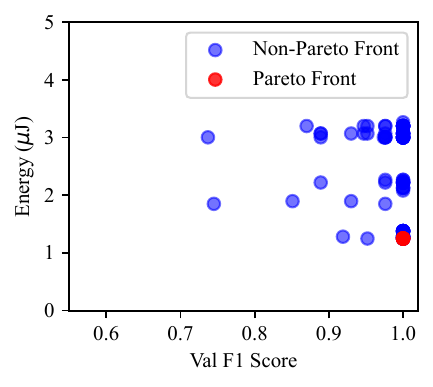}
        \caption{1D-CNN}
    \end{subfigure}
    \hfill
    \begin{subfigure}[b]{0.23\textwidth}
        \centering
        \includegraphics[width=\linewidth]{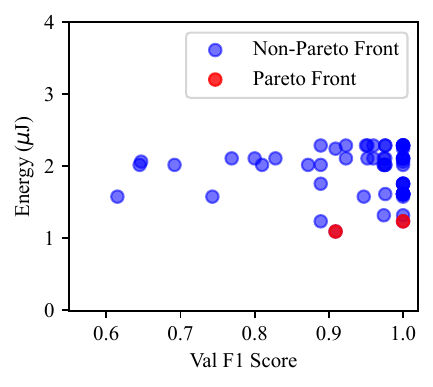}
        \caption{1D-SepCNN}
    \end{subfigure}
    \hfill
    \begin{subfigure}[b]{0.24\textwidth}
        \centering
        \includegraphics[width=\linewidth]{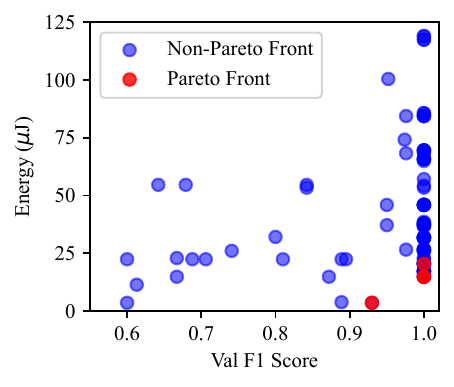}
        \caption{LSTM}
    \end{subfigure}
    \hfill
    \begin{subfigure}[b]{0.24\textwidth}
        \centering
        \includegraphics[width=\linewidth]{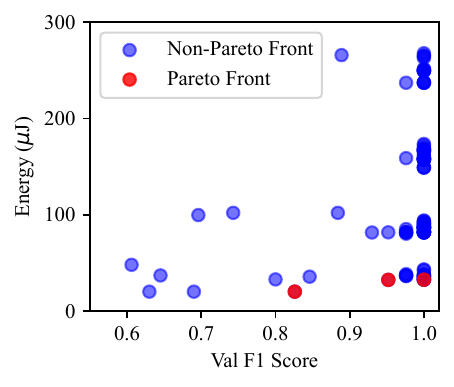}
        \caption{Transformer}
    \end{subfigure}
    \caption{Validation F1 score versus energy consumption for deployable configurations of four models on the XC7S15 FPGA. Each dot represents a configuration satisfying all hardware constraints, with Pareto front highlighted in red.}
    \label{fig:pareto_all}
\end{figure*}
\begin{table*}[!htpb]
\centering
\caption{Selected model configurations with highest test F1 scores under integer-only inference on the XC7S15 FPGA.}
\label{tab:exp1}
\resizebox{\textwidth}{!}{ 
\renewcommand{\arraystretch}{1.15}
\begin{tabular}{|c|c|c|c|c|c|c|c|c|c|c|c|c|c|}
\hline

\multirow{2}{*}{Model} & \multicolumn{4}{c|}{Configuration} & \multirow{2}{*}{Params} & \multicolumn{2}{c|}{Test F1 score} & \multirow{2}{*}{\begin{tabular}[c]{@{}c@{}}LUTs\\ (\%)\end{tabular}} & \multirow{2}{*}{\begin{tabular}[c]{@{}c@{}}BRAMs\\ (\%)\end{tabular}} & \multirow{2}{*}{\begin{tabular}[c]{@{}c@{}}DSPs\\ (\%)\end{tabular}} & \multirow{2}{*}{\begin{tabular}[c]{@{}c@{}}Energy\\ ($\mu$J)\end{tabular}} & \multirow{2}{*}{\begin{tabular}[c]{@{}c@{}}Power$^{*}$\\ (mW)\end{tabular}} & \multirow{2}{*}{\begin{tabular}[c]{@{}c@{}}Latency$^{**}$\\ (ms)\end{tabular}} \\ \cline{2-5} \cline{7-8}

& b & bs & lr ($\times10^{-4})$ & variable$^{\dagger}$ & & \multicolumn{1}{c|}{FP32} & Quantized  &  & & & & & \\ \hline\hline

1D-CNN & 4 & 48 & 3.367 & 3 & 173 & 0.889 & 0.900 ($\uparrow$1.24\%) & 15.80 & 0.0 & 30.0 & 1.247 & 39.0 & 0.032 \\ \hline

1D-SepCNN & 6 & 24 & 5.274 & 3  & 137 & 0.894 & 0.831 ($\downarrow7.05\%$)& 26.05 & 0.0 & 45.0 & 1.235 & 44.0 & 0.028\\ \hline 

LSTM & 8 & 32 & 3.627 & 24 & 2,738 & 0.911 & 0.889 ($\downarrow$2.47\%)& 31.20 & 25.0 & 55.0 & 20.318 & 59.0 & 0.344 \\ \hline 

Transformer & 4 & 16 & 4.325  & 8 & 922 &  0.916 & 0.937 ($\uparrow$2.29\%) & 35.21 & 85.0 & 65.0 & 32.314 & 60.0 & 0.539 \\ \hline 

\multicolumn{14}{l}{$b=$quantization bitwidth, $bs=$batch size, $lr=$learning rate, Params = number of model parameters.} \\
\multicolumn{14}{l}{$^{\dagger}$Model-specific variable: $\text{num}_\text{blocks}$ for 1D-CNN and 1D-SepCNN, $h_\text{size}$ for LSTM (hidden size), and $d_\text{model}$ for Transformer (embedding dimension).} \\
\multicolumn{14}{l}{$^{*}$Power estimated from Vivado synthesis reports was validated on the actual FPGA hardware at 28.0\textdegree C, showing a deviation within 5.4\%.} \\
\multicolumn{14}{l}{$^{**}$Latency measured on actual hardware deviates by 1.7\% from simulation results.} 
\end{tabular}
\vspace{-25pt}
}
\end{table*}

Figure~\ref{fig:pareto_all} visualizes the deployable model configurations of four model types regarding validation F1 score and energy consumption. As shown in Figure~\ref{fig:pareto_all}(a), 1D-CNN achieves consistently high efficiency, with F1 scores above 0.737 and energy consumption between 1.247 and 3.265 $\mu$J. In Figure~\ref{fig:pareto_all}(b), 1D-SepCNN yields F1 scores above 0.616 and energy between 1.094 and 2.288 $\mu$J. LSTM (see Figure~\ref{fig:pareto_all}(c)) shows greater variability, with F1 scores above 0.60 and energy consumption ranging from 3.522 to 118.98 $\mu$J. Transformer (see Figure~\ref{fig:pareto_all}(d)) spans the broadest trade-off range, with F1 scores from 0.606 to 1.0 and energy ranging from 20.067 to 267.818 $\mu$J.
Among these deployable configurations, the red markers indicate the Pareto fronts that achieve the best possible trade-offs between F1 score and energy consumption. Notably, the Pareto front of 1D-CNN forms a tight cluster in the high-F1 score, low-energy region, whereas the Pareto fronts of LSTM and Transformer exhibit wider spreads, indicating potential for a higher F1 score but at increased energy.

We further identify the best-performing configuration from each model’s Pareto front based on test F1 scores under integer-only inference, as summarized in Table~\ref{tab:exp1}. All reported power and latency values were cross-validated on the actual FPGA hardware. Specifically, the measured power at 28.0\textdegree C deviates by no more than 5.4\% from Vivado synthesis estimates, while the latency differs by 1.7\% from simulation results. 

All selected configurations achieve strong predictive performance (F1 score~$\geq$~0.831), even when operating under aggressive quantization and strict energy constraints.
With only 173 parameters, the selected 4-bit quantized 1D-CNN model achieves a quantized F1 score of 0.900, even an improvement of 1.24\% over its FP32 counterpart. It requires the fewest hardware resources (15.8\% LUTs, 0.0\% BRAMs, and 30\% DSPs) and operates at low power (39 mW). The resulting energy consumption is just 1.247 $\mu$J per inference, with a latency of 0.032 ms.
In contrast, the 1D-SepCNN model has an even smaller parameter count (137) and achieves a slightly higher F1 score in FP32 precision (0.894). Despite its architectural efficiency in reducing both parameter count and computation, it exhibits greater sensitivity to quantization with the selected model configuration, showing a substantial 7.05\% drop in F1 score (from 0.894 to 0.831), even under moderate 6-bit quantization. Its deployment requires moderately more hardware resources (26.05\% LUTs and 45\% DSPs) and 12.82\% higher power consumption (44 mW vs. 39 mW). Nevertheless, it maintains the lowest inference latency (0.028 ms), resulting in a competitive energy cost of 1.235 $\mu$J.

The chosen 8-bit quantized LSTM model contains substantially more parameters (2,738) than the other models. It achieves a quantized F1 score of 0.889, with a moderate 2.47\% drop from its FP32 baseline. However, it requires higher deployment cost: 31.2\% LUTs, 25.0\% BRAMs, and 55\% DSPs. At 59 mW power consumption and 0.344 ms latency due to its inherently sequential architecture, the resulting energy per inference reaches 20.318 $\mu$J, over 16$\times$ that of the 1D-CNN. In addition, LSTM suffers from slow training convergence and compilation. On average, a full end-to-end deployment takes 14.36 minutes, substantially longer than 1D-CNN (4.56 minutes), 1D-SepCNN (5.07 minutes), and Transformer (7.46 minutes). These findings underscore the relatively high training cost of LSTM, which slows down the deployment pipeline.

Moreover, the obtained Transformer model, with 922 parameters and an embedding dimension of 8, achieves the highest quantized test F1 score (0.937), outperforming its FP32 counterpart by 2.29\%. However, this gain comes at the cost of increased complexity. Power consumption is also highest at 60 mW, with a latency of 0.539 ms, resulting in an energy cost of 32.314 $\mu$J, more than 26$\times$ that of the 1D-CNN. 

Overall, all four models satisfy the deployment constraints after quantization, but their efficiency profiles vary markedly. CNN-based models demonstrate the most favorable trade-offs across F1 score and energy usage, making them highly suitable for real-time, low-power applications. In contrast, LSTM and Transformer models offer stronger representational capacity and comparative or higher test F1 score but incur higher resource usage and energy cost. However, even the most resource-intensive Transformer model completes inference within 0.539 ms, well below the 125 ms latency bound for real-time feedback.

\subsection{Stage 2: Cross-FPGA Validation}
\label{subsec:exp2}

To evaluate cross-platform portability, we deploy the best-performing configurations from Table~\ref{tab:exp1} onto a more constrained FPGA platform: the Lattice iCE40UP5K. Compared to the AMD XC7S15 (8,000 6-LUTs, 20 DSPs, and 10 BRAMs), the iCE40UP5K offers significantly fewer resources (5,280 4-LUTs, 8 DSPs, and 30 ERBs), imposing tighter constraints on model size and complexity.

We first examine the deployability of the LSTM model. The selected configuration in Table~\ref{tab:exp1} exceeded the iCE40UP5K’s LUT budget, reaching 103\% utilization and thus failing to deploy. To explore its feasibility under stricter resource limits, we conducted an additional Optuna-based search tailored to the iCE40UP5K at a reduced clock frequency of 20 MHz. This search yielded a Pareto-optimal configuration with a hidden size of 24, batch size of 40, and learning rate of 8.796$\times10^{-4}$. Under 6-bit quantization, the resulting model became deployable and significantly reduced energy consumption (20.318 $\mu$J vs 4.408 $\mu$J). However, this gain came at the cost of a 2.52\% drop in quantized F1 score and a fivefold increase in inference latency (from 0.344 ms to 1.722 ms).
In contrast, no deployable configuration was found for the Transformer model under the same resource constraints. Even after increasing the downsampling factor to shorten input sequences, the model required overly aggressive quantization (e.g., 4-bit) to fit within the LUT and DSP limits, resulting in unacceptable accuracy degradation. 

\begin{table}[!htpb]

\centering
\caption{Deployment performance on iCE40UP5K FPGA using the selected model configurations from Table \ref{tab:exp1}}
\label{tab:exp2}
\resizebox{\columnwidth}{!}{ 
\renewcommand{\arraystretch}{1.15}
\begin{tabular}{|c|c|c|c|c|c|c|}
\hline

\multirow{2}{*}{Model} & \multirow{2}{*}{\begin{tabular}[c]{@{}c@{}}LUTs\\ (\%)\end{tabular}} & \multirow{2}{*}{\begin{tabular}[c]{@{}c@{}}BRAMs\\ (\%)\end{tabular}} & \multirow{2}{*}{\begin{tabular}[c]{@{}c@{}}DSPs\\ (\%)\end{tabular}} & \multirow{2}{*}{\begin{tabular}[c]{@{}c@{}}Energy\\ ($\mu$J)\end{tabular}} & \multirow{2}{*}{\begin{tabular}[c]{@{}c@{}}Power$^{*}$\\ (mW)\end{tabular}} & \multirow{2}{*}{\begin{tabular}[c]{@{}c@{}}Latency$^{**}$\\ (ms)\end{tabular}} \\ 
&  & & & & & \\ \hline\hline

1D-CNN & 42.99 & 43.33 & 100.0 & 0.366 & 2.290 & 0.160   \\ \hline
1D-SepCNN & 80.21 & 46.67 & 100.0 & 0.350 & 2.494 & 0.140 \\ \hline 

\multicolumn{7}{l}{$^{*}$Measured on hardware at 28.0\textdegree C, power estimates deviated by 6.3\%.} \\
\multicolumn{7}{l}{$^{**}$Latency measured on actual hardware deviates by 1.5\% from estimates.} \\

\end{tabular}
\vspace{-5pt}
}
\end{table}

Both CNN-based models are successfully deployed on the iCE40UP5K FPGA running at 20 MHz, as summarized in Table~\ref{tab:exp2}. For the 1D-CNN model, latency increases from 0.032 ms (on XC7S15) to 0.160 ms due to the reduced clock frequency. However, its power consumption drops sharply from 39 mW to just 2.290 mW (over 17$\times$ lower) thanks to the iCE40UP5K’s ultra-low static power. As a result, energy per inference decreases from 1.247~$\mu$J to 0.366~$\mu$J, yielding a 3.41$\times$ reduction.
The 1D-SepCNN model shows slightly higher power consumption (2.494 mW), but achieves faster inference (0.140 ms), leading to the lowest energy cost of 0.350~$\mu$J per inference. Based on the 1D-SepCNN deployment, we estimate battery lifetime under realistic operating conditions. Assuming the MCU remains in sleep mode (1.25 mW) and the IMU continuously streams data into its internal FIFO (2.28 mW), the FPGA performs 8 inferences per second, averaging only 0.0028 mW in compute power. Under these conditions, the system can sustain up to 13.6 days of continuous operation on a 320 mAh battery. 

These findings highlight the strong cross-platform deployability of CNN-based models, while underscoring that LSTM and Transformer models will require further architectural simplification or software-hardware co-design strategies to become feasible on ultra-low-power FPGAs. 

\subsection{Stage 3: Cross-Participant Generalization Validation}
\label{subsec:exp3}

\begin{figure*}[!htbp]
    \centering
    \includegraphics[width=\textwidth]{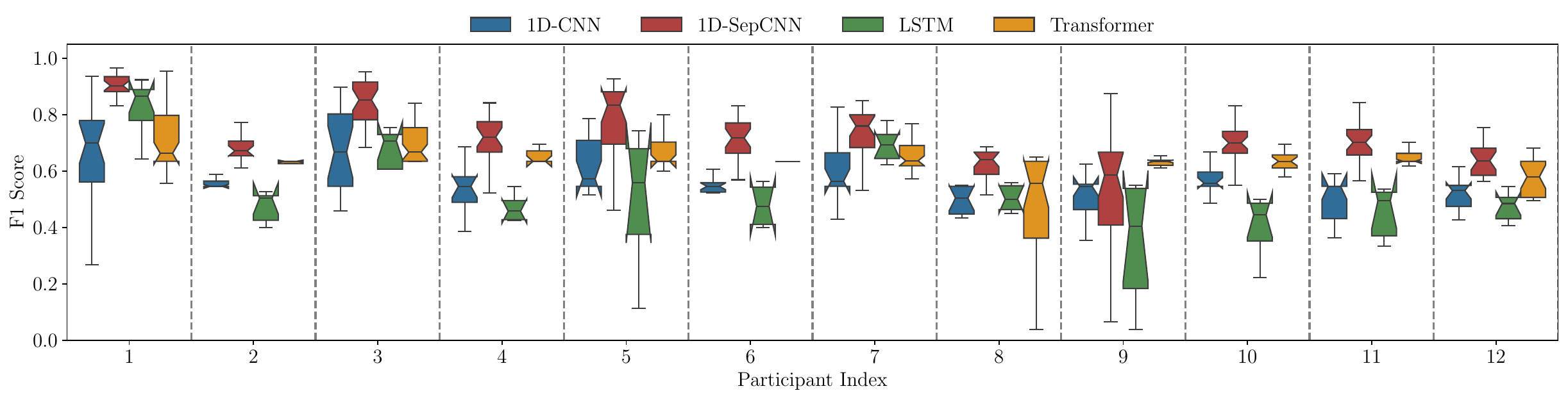}
    \caption{Cross-participant distribution of quantized test F1 Scores for each selected Model configuration from Table \ref{tab:exp1}}
    \label{fig:exp3} 
\vspace{-10pt}
\end{figure*}

At the final stage, we apply all model configurations obtained from Participant~1 to all 12 participants to evaluate the generalization of model configurations identified in Stage~1 (see Table~\ref{tab:exp1}). For each participant, each model is retrained from scratch for 50 sessions.
Figure~\ref{fig:exp3} presents the distribution of test F1 scores of quantized models across 12 participants. Notched boxplots are used to visualize both median values and their confidence intervals.

Regarding Participant~1, we observe a notable change in model rankings of predictive performance compared to Table~\ref{tab:exp1}. Although 1D-SepCNN exhibited the lowest quantized test F1 score in Stage~1, it achieves the highest median and exhibits low variability in Stage~3. In contrast, 1D-CNN, which previously performed well, shows significantly lower median performance and higher variance.
These observations reveal the limitation of the Optuna-guided QAT. Although QAT incorporates quantization error during training, the search process optimizes for the validation F1 score obtained through fake quantization, which does not always reflect the actual performance under integer-only inference on the test set. This mismatch stems from differences between the validation and test data distributions, which can lead to suboptimal quantization parameters (e.g., scaling factors) for testing.
Furthermore, each candidate configuration is trained only once during the Optuna search, leaving outcomes vulnerable to stochastic variation. Certain configurations may appear better or worse purely due to chance. These insights underscore the importance of incorporating robustness-aware objectives (such as retraining-averaged performance) into the search process.

Extending the analysis across all participants reveals consistent trends. 1D-SepCNN consistently delivers the highest average performance (0.847) with narrow confidence intervals on most participants, highlighting its strong generalization capabilities. Transformer shows competitive median accuracy and frequently ranks in the top two, but exhibits higher inter-user variance. LSTM demonstrates broader variability and generally ranks lower, while 1D-CNN shows the weakest performance with both low median F1 scores and wide dispersion.

Compared to Joo et al.~\cite{Joo2022} (see Table~\ref{tab:imu_comparison}), who achieved 0.98 F1 score using 1D-CNNs on wrist-mounted IMU data across 17 participants, our best model reports a lower average F1 score (0.847 vs. 0.98). However, their evaluation is conducted on high-performance platforms and omits deployment constraints such as energy consumption and hardware feasibility. In contrast, our models are deployed on resource-constrained embedded FPGAs, reflecting an intentional trade-off between accuracy and deployability. Rather than maximizing accuracy alone, we select configurations that balance accuracy and energy efficiency. Notably, when energy constraints are relaxed, our quantized models can reach an F1 score of 0.990 with an energy cost of 1.375~$\mu$J. Moreover, while Joo et al. report 27.8 ms latency for 1D-CNN inference on a batch of 100 samples (i.e., 0.28 ms per sample), our 1D-CNN achieves a comparable latency of 0.140~ms for a single input window with a length of 25 on the ultra-low-power Lattice iCE40UP5K. These results highlight the competitiveness of our approach under real-world deployment constraints, supporting its applicability to long-term, autonomous wearable use.

\section{Conclusion and Future Work}
\label{sec:conclusion_future_work}

This study presents \textit{StrikeWatch}, a fully self-contained wrist-worn system for real-time gait recognition in outdoor running scenarios. Addressing the low-fidelity wrist IMU signals and limited compute budgets, we integrated four compact DL models into a software-hardware co-designed pipeline for deployment on embedded FPGAs. Using a custom prototype and a real-world dataset collected from 12 participants, we systematically explored the trade-offs between model complexity and hardware efficiency across two representative FPGAs. Our results demonstrate that accurate, low-latency, and energy-efficient footstrike classification can be achieved entirely on-device. Notably, 1D-SepCNN achieves the best overall balance between accuracy and energy consumption, supporting over 13.6-day operation on a LiPo battery. These results provide actionable insights for designing wearable systems capable of running activity recognition in real-world settings.

In the future, we plan to expand the dataset by including more participants with diverse footwear and terrains to address the observed variability in cross-user performance. We also aim to improve the robustness of quantized models by exploring retraining strategies that reduce sensitivity to data distribution, which affects deployment consistency in our current pipeline. 

\bibliographystyle{IEEEtran}
\bibliography{reference}

@inproceedings{ling2024aiot,
  title={Integer-only Quantized {Transformers} for Embedded {FPGA}-based Time-series Forecasting in {AIoT}},
  author={Ling, Tianheng and Qian, Chao and Schiele, Gregor},
  booktitle={IEEE Annual Congress on Artificial Intelligence of Things},
  year={2024},
  organization={IEEE}
}

@inproceedings{becnel2022tiny,
  title={{Tiny time-series {Transformers}: {R}ealtime multi-target sensor inference at the edge}},
  author={Becnel, Tom and Kelly, Kerry and Gaillardon, Pierre-Emmanuel},
  booktitle={International Conference on Omni-layer Intelligent Systems},
  year={2022},
  organization={IEEE}
}

@inproceedings{muhamad2023design,
  title={Design and Implementation of Wearable {IMU} Sensor System for Heel-Strike and Toe-Off Gait Parameter Measurement},
  author={Muhamad, MF and Razak, AHA and Halim, AK and Idros, MF Md and Osman, FN and Al Junid, SAM and Chee, S Pawan},
  booktitle={International Conference on Applied Electronics and Engineering},
  pages={1--5},
  organization={IEEE},
  year={2023},
}

@inproceedings{Yuwono2013,
  title={Unsupervised segmentation of heel-strike {IMU} data using rapid cluster estimation of wavelet features},
  author={Yuwono, Mitchell and Su, Steven W and Moulton, Bruce D and Nguyen, Hung T},
  booktitle={35th Annual International Conference of the IEEE Engineering in Medicine and Biology Society},
  year={2013},
  organization={IEEE}
}

@inproceedings{karakasis2021f,
  title={{F-VESPA}: A kinematic-based algorithm for real-time heel-strike detection during walking},
  author={Karakasis, Chrysostomos and Artemiadis, Panagiotis},
  booktitle={International Conference on Intelligent Robots and Systems},
  year={2021},
  organization={IEEE}
}

@article{Young2023,
  title={Bespoke fuzzy logic design to automate a better understanding of running gait analysis},
  author={Young, Fraser and Stuart, Samuel and McNicol, Robert and Morris, Rosie and Downs, Craig and Coleman, Martin and Godfrey, Alan},
  journal={Journal of Biomedical and Health Informatics},
  year={2022},
  volume={27},
  number={5},
  pages={2178--2185},
  publisher={IEEE}
}

@article{Giandolini2014,
  title={A simple field method to identify foot strike pattern during running},
  author={Giandolini, Marl{\`e}ne and Poupard, Thibaut and Gimenez, Philippe and Horvais, Nicolas and Millet, Guillaume Y and Morin, Jean-Beno{\^\i}t and Samozino, Pierre},
  journal={Journal of biomechanics},
  pages={1588--1593},
  year={2014},
  publisher={Elsevier}
}

@article{Mahoney2024,
  title={Identification of footstrike pattern using accelerometry and {Machine Learning}},
  author={Mahoney, Joseph M and Rhudy, Matthew B and Outerleys, Jereme and Davis, Irene S and Altman-Singles, Allison R},
  journal={Journal of Biomechanics},
  year={2024},
  volume={174},
  publisher={Elsevier}
}

@inproceedings{akiba2019optuna,
  title={Optuna: A next-generation hyperparameter optimization framework},
  author={Akiba, Takuya and Sano, Shotaro and Yanase, Toshihiko and Ohta, Takeru and Koyama, Masanori},
  booktitle={Proceedings of the 25th ACM SIGKDD international conference on knowledge discovery \& data mining},
  pages={2623--2631},
  year={2019}
}

@article{hassan2017footstriker,
  title={Footstriker: An {EMS}-based foot strike assistant for running},
  author={Hassan, Mahmoud and Daiber, Florian and Wiehr, Frederik and Kosmalla, Felix and Kr{\"u}ger, Antonio},
  journal={Proceedings of the ACM on Interactive, Mobile, Wearable and Ubiquitous Technologies},
  volume={1},
  number={1},
  pages={1--18},
  year={2017},
  publisher={ACM New York, NY, USA}
}

@inproceedings{schiewe2020study,
  title={A study on real-time visualizations during sports activities on smartwatches},
  author={Schiewe, Alexander and Krekhov, Andrey and Kerber, Frederic and Daiber, Florian and Kr{\"u}ger, Jens},
  booktitle={Proceedings of the 19th International Conference on Mobile and Ubiquitous Multimedia},
  pages={18--31},
  year={2020}
}

@inproceedings{Kandpal2023,
  title={Human activity recognition in smart cities from smart watch data using {LSTM} {Recurrent Neural Networks}},
  author={Kandpal, Meenakshi and Sharma, Bhisham and Barik, Rabindra K and Chowdhury, Subrata and Patra, Sudhansu Shekhar and Dhaou, Imed Ben},
  booktitle={1st International Conference on Advanced Innovations in Smart Cities},
  year={2023},
  organization={IEEE}
}

@misc{RunningImage,
  title = {Foot Strike Position \& Their Effects in Running: Part 2},
  howpublished = {\url{https://www.healthystep.co.uk/advice/foot-strike-position-affects-part-2/}},
  author = {HealthyStep},
  year = {2025},
  note = {Accessed: 2025-05-29}
}

@article{burke2021risk,
  title={Risk factors for injuries in runners: {A} systematic review of foot strike technique and its classification at impact},
  author={Burke, Aoife and Dillon, Sarah and O’Connor, Siobh{\'a}n and Whyte, Enda F and Gore, Shane and Moran, Kieran A},
  journal={Orthopaedic journal of sports medicine},
  volume={9},
  number={9},
  year={2021},
  publisher={Sage Publications Sage CA: Los Angeles, CA}
}

@inproceedings{cola2017personalized,
  title={Personalized gait detection using a wrist-worn accelerometer},
  author={Cola, Guglielmo and Avvenuti, Marco and Musso, Fabio and Vecchio, Alessio},
  booktitle={International Conference on Wearable and Implantable Body Sensor Networks},
  year={2017},
  organization={IEEE}
}

@article{ling2025automated,
  title={Automated Energy-Aware Time-Series Model Deployment on Embedded {FPGAs} for Resilient Combined Sewer Overflow Management},
  author={Ling, Tianheng and Singh, Vipin and Qian, Chao and Biessmann, Felix and Schiele, Gregor},
  journal={arXiv preprint arXiv:2508.13905},
  year={2025}
}

@article{Joo2022,
  title={Estimation of fine-grained foot strike patterns with wearable smartwatch devices},
  author={Joo, Hyeyeoun and Kim, Hyejoo and Ryu, Jeh-Kwang and Ryu, Semin and Lee, Kyoung-Min and Kim, Seung-Chan},
  journal={International journal of environmental research and public health},
  volume={19},
  number={3},
  pages={1279},
  year={2022},
  publisher={MDPI}
}

@article{benson2022real,
  title={Is this the real life, or is this just laboratory? {A} scoping review of {IMU}-based running gait analysis},
  author={Benson, Lauren C and R{\"a}is{\"a}nen, Anu M and Clermont, Christian A and Ferber, Reed},
  journal={Sensors},
  volume={22},
  number={5},
  pages={1722},
  year={2022},
  publisher={MDPI}
}

@article{Zago2021,
  title={{Machine Learning}-based determination of gait events from foot-mounted inertial units},
  author={Zago, Matteo and Tarabini, Marco and Delfino Spiga, Martina and Ferrario, Cristina and Bertozzi, Filippo and Sforza, Chiarella and Galli, Manuela},
  journal={Sensors},
  pages={839},
  year={2021},
  volume={21},
  publisher={MDPI}
}

@inproceedings{roggen2022wearable,
  title={Wearable {FPGA} platform for accelerated {DSP} and {AI} applications},
  author={Roggen, Daniel and Cobden, Robert and Pouryazdan, Arash and Zeeshan, Muhammad},
  booktitle={International Conference on Pervasive Computing and Communications Workshops and other Affiliated Events},
  pages={66--69},
  year={2022},
  organization={IEEE}
}

@inproceedings{qian2023elasticai,
  title={{ElasticAI}: Creating and deploying energy-efficient {Deep Learning} accelerator for pervasive computing},
  author={Qian, Chao and Ling, Tianheng and Schiele, Gregor},
  booktitle={International Conference on Pervasive Computing and Communications Workshops and other Affiliated Events},
  pages={297--299},
  year={2023},
  organization={IEEE}
}

@article{li2024hardware,
  title={A Hardware and Software Co-Design for Energy-Efficient {Neural Network Accelerator} With Multiplication-Less Folded-Accumulative {PE} for Radar-Based Hand Gesture Recognition},
  author={Li, Fan and Guan, Yunqi and Ye, Wenbin},
  journal={IEEE Transactions on Very Large Scale Integration (VLSI) Systems},
  year={2024},
  publisher={IEEE}
}

@article{chen2024eciton,
  title={Eciton: Very low-power {Recurrent Neural Network} accelerator for real-time inference at the edge},
  author={Chen, Jeffrey and Jun, Sang-Woo and Hong, Sehwan and He, Warrick and Moon, Jinyeong},
  journal={ACM Transactions on Reconfigurable Technology and Systems},
  volume={17},
  number={1},
  pages={1--25},
  year={2024},
  publisher={ACM New York, NY}
}

@article{chen2021mlof,
  title={{MLoF}: {Machine Learning} accelerators for the low-cost {FPGA} platforms},
  author={Chen, Ruiqi and Wu, Tianyu and Zheng, Yuchen and Ling, Ming},
  journal={Applied sciences},
  year={2021},
  publisher={MDPI},
  volume={12},
  number={1},
  pages={89},
}

@article{lingaiot2025smatbale,
   title={Enabling Vibration-Based Gesture Recognition on Everyday Furniture via Energy-Efficient {FPGA} Implementation of {1D Convolutional Networks}},
  author={Shibata, Koki and Ling, Tianheng and Qian, Chao and Matsui, Tomokazu and Suwa, Hirohiko and Yasumoto, Keiichi and Schiele, Gregor},
  journal={IEEE Annual Congress on Artificial Intelligence of Things (AIoT)},
  year={2025},
  organization={IEEE}
}

@INPROCEEDINGS{ling2025deployment,
  author={Ling, Tianheng and Qian, Chao and Haßler, Lukas Johannes and Schiele, Gregor},
  booktitle={2025 IEEE Computer Society Annual Symposium on VLSI (ISVLSI)}, 
  title={Automating Versatile Time-Series Analysis with Tiny {Transformers} on Embedded {FPGAs}}, 
  year={2025},
  volume={1},
  number={},
  pages={1-6},
  doi={10.1109/ISVLSI65124.2025.11130202}}
\end{document}